\documentclass[fleqn]{an}
\usepackage{times}
\usepackage{graphicx}
\usepackage{color}
\usepackage{natbib}
\bibpunct{(}{)}{;}{a}{}{,}

\usepackage{amsmath}
\usepackage{amssymb}
\def\bmath#1{\boldsymbol{#1}}

\def\secref#1{Sect.~\ref{#1}}
\def\Secref#1{Section~\ref{#1}}

\def\eqref#1{Eq.~(\ref{#1})}

\def\eqsand#1#2{Eqs.~(\ref{#1}) and~(\ref{#2})}

\def\figref#1{Fig.~\ref{#1}}
\def\Figref#1{Figure~\ref{#1}}

\def\tabref#1{Table~\ref{#1}}

\def\const{\mathrm{const}}

\def\bea{\begin{eqnarray}}
\def\eea{\end{eqnarray}}

\def\be{\begin{equation}}
\def\ee{\end{equation}}

\def\({\left(}
\def\){\right)}
\def\<{\langle}
\def\>{\rangle}

\def\diff{\mathrm{d}}
\def\dd{\partial}
\def\vdel{\bmath{\nabla}}

\def\vx{\bmath{\hat x}}
\def\vy{\bmath{\hat y}}
\def\vz{\bmath{\hat z}}
\def\vu{\bmath{u}}
\def\vB{\bmath{B}}
\def\vf{\bmath{f}}

\def\Rm{\mathrm{Rm}}
\def\Rmc{\Rm_{\mathrm{c,fd}}}
\def\Re{\mathrm{Re}}
\def\Pm{\mathrm{Pm}}

\def\eps{\epsilon}
\def\urms{u_{\mathrm{rms}}}
\def\Brms{B_{\mathrm{rms}}}

\def\xyavg#1{\left\langle{#1}\right\rangle_{xy}}
\def\tavg#1{\left\langle{#1}\right\rangle_{t}}
\def\tavg#1{\overline{#1}}

\def\curl{\vdel\times}

\def\cite#1{}

\def\sm#1{ {\scriptscriptstyle #1}   }
\def\Bm{B^{\sm<}} 
\def\vBm{\bmath{\Bm}}
\def\lb{l_{B}}

\graphicspath{figs}

\begin{document}

\title{Numerical experiments on dynamo action in sheared and rotating\\
 turbulence}

%\pagewiselinenumbers
%\renewcommand\linenumberfont{\normalfont\tiny\color{blue}}
%\setlength\linenumbersep{1mm}

\author{%
T.~A.~Yousef\inst{1,2}\fnmsep\thanks{Corresponding author: tarek@pvv.org} \and
T.~Heinemann\inst{1}\and
F.~Rincon\inst{3}\and
A.~A.~Schekochihin\inst{2}\and
N.~Kleeorin\inst{4}\and
I.~Rogachevskii\inst{4}\and
S.~C.~Cowley\inst{2,5}\and
J.~C.~McWilliams\inst{6}
}
\institute{
Department of Applied Mathematics and Theoretical Physics, University of
Cambridge, Cambridge CB3 0WA, UK 
\and
Plasma Physics Group, Blackett Laboratory, Imperial College, London SW7 2AZ, UK 
\and
Laboratoire d'Astrophysique de Toulouse-Tarbes, Universit\'e de Toulouse, CNRS,
14 avenue Edouard Belin, F-31400 Toulouse, France 
\and
Department of Mechanical Engineering, 
The Ben-Gurion University of the Negev, P.~O.~Box 653, Beer-Sheva 84105, Israel
\and
Euratom/UKAEA Fusion Association, Culham Science Centre, Abington OX14 3DB, UK
\and
Department of Atmospheric Sciences, University of California, Los Angeles, California 90095-1565, USA}

%\date{Received; accepted; published online}
\date{\today; \quad e-print {\tt arxiv 0000000000}}

\Pagespan{737}{}
\Yearpublication{2008}%
\Yearsubmission{2008}%
\Month{6}%   
\Volume{329}%  
\Issue{7}% 
\DOI{10.1002/asna.200811018} 

\received{2008 Jun 28}
\accepted{2008 Jul 3}
\publonline{2008 Aug 30}

\abstract{Numerical simulations of forced turbulence in elongated shearing boxes
are carried out to demonstrate that a nonhelical turbulence in conjunction 
with a linear shear can give rise to a mean-field dynamo. 
Exponential growth of magnetic field at scales larger than the outer 
(forcing) scale of the turbulence is found. 
Over a range of values of the shearing rate $S$ spanning approximately 
two orders of magnitude, the growth rate of the magnetic 
field is proportional to the imposed shear, $\gamma\propto S$, 
while the characteristic spatial scale of the field is $\lb\propto S^{-1/2}$. 
The effect is quite general: 
earlier results for the nonrotating case by \cite{}Yousef et~al. (2008) 
are extended to shearing boxes with Keplerian rotation; 
it is also shown that the shear dynamo mechanism operates both below and above 
the threshold for the fluctuation dynamo. 
The apparently generic nature of the shear dynamo effect makes it an attractive 
object of study for the purpose of understanding the generation of magnetic fields 
in astrophysical systems.} 

%\keywords{MHD---shear---turbulence---magnetic fields---dynamo}

\maketitle

\def\runtable{
\renewcommand{\thefootnote}{\alph{footnote}}
\def\hlf{$1/2$}
\def\qrt{$1/4$}
\def\eth{$1/8$}
\def\fna{\footnotemark[1]}
\def\fnb{\footnotemark[2]}
\def\fnc{\footnotemark[3]}

\begin{table*}[t!]
\caption{Index of runs.}
\begin{minipage}{\textwidth}
%\centering
\begin{tabular}{llrllllll}
\hline\noalign{\smallskip}%\hline
\label{tb:runs}
Run&~~~$S$&$\Omega$&$L_z$& Resolution&$\nu=\eta$& ~~$\gamma$&$\lb$&
$\tavg{\Bm_y/\Bm_x}$~\fna\\[2pt]
\hline\noalign{\smallskip}
S1  &  $-2$     &  0     & 8  & $32^2\times256$& $1\times10^{-2}$ & 0.0161  & 3.7  & 11.5 \\ 
S2  &  $-2$     &  0     & 16 & $32^2\times512$& $1\times10^{-2}$ & 0.021   & 3.8  & 11.1 \\ 
S3  &  $-1$     &  0     & 8  & $32^2\times256$& $1\times10^{-2}$ & 0.0030  & 4.6  & 10.0 \\ 
S4  &  $-1$     &  0     & 16 & $32^2\times512$& $1\times10^{-2}$ & 0.0124  & 5.4  & 10.2 \\ 
S5  &  $-1$     &  0     & 32 &$32^2\times1024$& $1\times10^{-2}$ & 0.0092  & 5.2  & 9.9  \\ 
S6  &  $-1$     &  0     & 64 &$32^2\times2048$& $1\times10^{-2}$ & 0.0121  & 5.1  & 9.5  \\ 
S7  &  $-1/2$   &  0     & 16 &$32^2\times512$ & $1\times10^{-2}$ & 0.0040  & 6.8  & 9.1  \\ 
S8  &  $-1/2$   &  0     & 32 &$32^2\times1024$& $1\times10^{-2}$ & 0.0058  & 7.1  & 9.0  \\ 
S9  &  $-1/2$   &  0     & 64 &$32^2\times2048$& $1\times10^{-2}$ & 0.0055  & 7.3  & 9.0  \\ 

S10 &  $-1/4$   &  0     & 64 &$32^2\times2048$& $1\times10^{-2}$ & 0.0025  & 9.7  & 8.3  \\ 
S11 &  $-1/4$   &  0     & 128&$32^2\times4096$& $1\times10^{-2}$ & 0.0025  & 9.9  & 8.2  \\ 
S12 &  $-1/8$   &  0     & 64 &$32^2\times2048$& $1\times10^{-2}$ & 0.00094 & 13.1 & 8.1  \\ 
S13 &  $-1/8$   &  0     & 128&$32^2\times4096$& $1\times10^{-2}$ & 0.00092 &
13.5 & 8.1  \\[2pt]
\hline\noalign{\smallskip}
K1  &  $-1/2$   &$1/3$  & 32 &$32^2\times1024$& $1\times10^{-2}$ & 0.0067   & 7.0  & 8.9  \\ 
K2  &  $-1$     &$2/3$  & 16 &$32^2\times512$ & $1\times10^{-2}$ & 0.0174   & 5.0  & 9.7  \\ 
K3  &  $-2$     &$4/3$  &  8 &$32^2\times256$ & $1\times10^{-2}$ & 0.038    & 3.5  & 10.7 \\ 
K4  &  $-4$     &$8/3$  &  8 &$32^2\times256$ & $1\times10^{-2}$ & 0.086    & 2.6  & 12.7 \\ 
K5  &  $-8$     &$16/3$ &  8 &$32^2\times256$ & $1\times10^{-2}$ & 0.118    & 2.2  & 20   \\ 
K6  &  $-16$    &$32/3$ &  8 &$32^2\times256$ & $1\times10^{-2}$ & 0.147    & 2.1  & 44   \\ 
K7  &  $-32$    &$64/3$ &  8 &$32^2\times256$ & $1\times10^{-2}$ & 0.078    & 2.3  & 137  \\ 
K8  &  $-64$    &$128/3$&  8 &$32^2\times256$ & $1\times10^{-2}$ & 0.036    &
2.9  & 520  \\[2pt]
\hline\noalign{\smallskip}
FD1 &  $-1$     &$2/3$  &  8 & $64^2\times1024$&$1\times10^{-3}$  & 0.32\fnb & 5.8\fnc& 16.4\fnc \\ 
FD2 &  0        &$0$    &  8 & $64^2\times1024$&$1\times10^{-3}$  & 0.29\fnb &
--   & --   \\[2pt] 
\hline%\hline
\end{tabular}
\footnotetext[1]{This is actually 
$\tavg{[\langle{\Bm_y}^2\rangle_z/\langle{\Bm_x}^2\rangle_z]^{1/2}}$.
The reported values differ from the corresponding numbers in
\cite{}Yousef et~al.~2008. where we computed $\tavg{\xyavg{B_y}/\xyavg{B_x}}$
(fields averaged over $x$ and $y$) instead of 
$\tavg{[\langle{\Bm_y}^2\rangle_z/\langle{\Bm_x}^2\rangle_z]^{1/2}}$
(as erroneously claimed there).} 
\footnotetext[2]{Calculated for the
kinematic regime of the fluctuation dynamo (where $\Brms/\urms$ $<$ $10^{-2})$.
Refer to \figref{fig:fd} (left panel) for time evolution of $\urms$ and $\Brms$
for runs FD1--FD2. } 
\footnotetext[3]{Calculated for period after saturation of the fluctuation dynamo  ($t>200$).}
\end{minipage}

\end{table*}

}

\section{Introduction}

Turbulence is generally considered to play a fundamental role in the
generation and maintenance of magnetic fields found 
in a wide range of astrophysical systems. 
It is a numerically well established property of turbulence to 
amplify magnetic fluctuations at the same or smaller scales 
than the scales of the turbulent motions \citep{meneguzzi81,sch04,haugen04,iskakov07,sch07}. 
This type of dynamo is known as small-scale, or fluctuation, dynamo.
It is believed to be a universal property of turbulent systems 
and, at least in the case of large magnetic Prandtl numbers, 
a simple theoretical picture exists of the field amplification 
via random stretching by turbulent motions 
(\citealt{moffatt64,zeldovich84}; see also \citealt{sch04,SC07}).  
It often turns out to be more difficult to either establish 
or explain the generation of magnetic fields at 
much larger scales than the turbulence scale. 
Such fields are observed or believed to exist in many astrophysical 
bodies (stars, galaxies, disks), so the question of their origin 
is an important theoretical challenge. 

Mechanisms for the generation of such large-scale fields are
known as large-scale, or mean-field, dynamos.
A motley of such mean-field dynamos has been studied in the literature
\citep[e.g.,][]{moffatt78,krause80,brand05review,raedler07}. 
The precise way in which they operate often seems to be 
system dependent and introducing ever more realistic features 
into one's theoretical model produces ever more complicated behaviour. 
While such modelling is necessary for quantitative understanding, 
it is quite interesting to ask what are the generic ingredients 
required to produce large-scale fields. One such ingredient 
appears to be the presence of net kinetic helicity in the system: in many mean-field dynamos, 
the key generation mechanism is the so-called $\alpha$ effect \citep{steenbeck66}, 
whereby an assembly of non-mirror-symmetric velocity fluctuations 
having a nonzero net helicity are responsible for magnetic-field generation. 
The existence of a mean-field dynamo in helically forced turbulence 
is well established numerically \citep{brand01,maron02,brand08}. 
However, the requirement of net helicity may 
somewhat limit the applicability of the $\alpha$ effect. 
It is also far from certain that the direct link 
between kinetic helicity and mean-field generation via the $\alpha$ effect 
found in the model case where the helicity is injected by the random 
forcing, carries over to the cases where the helicity arises naturally 
(e.g., in a rotating convective layer; see \citealt{cattaneo06}). 
Therefore, there is a strong motivation for seeking alternative 
mean-field dynamo processes driven by nonhelical turbulence. 

It is clear that a nonhelical turbulence by itself cannot make 
large-scale fields. In recent years, many authors have argued that 
large-scale magnetic fields can be generated by nonhelical velocity 
fluctuations when acted upon by a large-scale shear: theoretical 
paradigms put forward to support such a dynamo, which we refer to as the shear
dynamo, have included the shear-current effect \citep{RK03,RK04}, 
the stochastic $\alpha$ effect (\citealt{vishniac97,silantev00,fedotov03,fedotov06,proctor07,brand07}; 
see, however, \citealt{kleeorin08}), 
negative-diffusivity type theories \citep{urpin99a,urpin99b,ruediger01,urpin02,urpin06}, 
shear amplification of the fluctuation-dynamo-generated  small-scale 
fields \citep{blackman98}.
While there is not yet agreement between theoreticians about 
the validity or areas of practical applicability 
of these models, it is clear that they are addressing 
a fundamental issue. Indeed, shear is an extremely 
generic property of astrophysical systems, so 
the idea of a { shear dynamo} gives us a 
particularly attractive scenario for ubiquitous generation 
of large-scale magnetic fields.  

Until recently, a numerical demonstration of this type of dynamo remained 
elusive. Originally motivated by the predictions of \citet{RK03,RK04}, 
we have previously performed numerical simulations of nonhelical
turbulence with a superimposed linear shear and demonstrated the
existence of the shear dynamo \citep{yousef08}. Theoretical understanding
of these numerical results is still poor. More work 
and, we believe, more information gathered from numerical experiments 
are needed in order to make progress towards understanding 
the properties of this dynamo and the underlying physical 
processes that produce it. This paper, which is an extension of the
work by \citet{yousef08}, aims at presenting a collection of new numerical results
regarding the existence and behaviour of the shear dynamo in various
regimes. We focus on three different cases of astrophysical
interest, namely shear dynamo in the presence of forced nonhelical
turbulence and a linear velocity shear, shear dynamo in the presence of
forced nonhelical turbulence and Keplerian differential rotation, and
finally shear dynamo in the presence of forced nonhelical turbulence,
Keplerian differential rotation and a small-scale fluctuation 
dynamo. In order to study the effects of shear and rotation, we adopt
a local rotating shearing sheet model. This model and the
corresponding numerical set-up are presented in \secref{sec:model}. In
\secref{sec:linear}, we consider the case of linear shear without
rotation. Results for the Keplerian regime are presented in
\secref{sec:kepler}. \Secref{sec:fd} describes some preliminary results on 
the shear dynamo in the presence of small scale magnetic fluctuations generated
by the fluctuation dynamo. A short discussion concludes the paper (\secref{sec:conc}).

\section{Model and numerical set up}
\label{sec:model}

We consider differentially rotating flows that
can locally be described in terms of a background shear flow
$\bmath{U} =Sx\vy$ rotating uniformly with a rotation rate $\Omega\vz$. 
In the framework of incompressible
magnetohydrodynamics (MHD), solenoidal velocity and magnetic
field perturbations to this flow evolve according to the following
equations written in the rotating frame:
\bea   
\label{eq:ns}
\lefteqn{\frac{\diff\vu}{\diff t} = 
-u_x(S+2\Omega)\vy+2\Omega u_y\vx-\frac{\vdel p}\rho}  \\
\nonumber 
&& \ \ \ +\ {\vB\cdot\vdel\vB} + \nu\nabla^2\vu + \vf, \\
\label{eq:induction}
\lefteqn{\frac{\diff\vB}{\diff t} = 
B_xS\vy+ \vB\cdot\vdel\vu +\eta\nabla^2\vB,} 
\eea
where $\diff/\diff t=\dd/\dd t + (\bmath{U}+\vu)\cdot\vdel$, $\vu$ is
the velocity deviation from the background flow ${\bmath U}$, $\vB$ is the
magnetic field normalised by $\sqrt{4\pi\rho}$, 
$\rho=1$ is the density, $p$ is the pressure determined by requiring
$\vdel\cdot\vu=0$, and $\nu$ and $\eta$ are the kinematic viscosity
and magnetic diffusivity coefficients, respectively. 

Using this local formulation allows us to study 
different instances of differentially rotating background
flows by changing the values of the shearing
and rotation rates $S$ and $\Omega$. In this paper, we consider two important cases: 
(a) shear with no rotation, $S\neq0$, $\Omega=0$ (see \secref{sec:linear})
and (b) Keplerian rotation, $\Omega=-2S/3$ (see \secref{sec:kepler}). 
The nonrotating case has been discussed by us previously 
\citep{yousef08} and is relevant to nonrotating systems as well as systems
where one expects the shear to be stronger than the rotation 
(one good astrophysical example is irregular galaxies, where 
large-scale magnetic fields are found in the absence of strong 
overall rotation;\footnote{We thank 
D.~Sokoloff for pointing this out to us.} see \citealt{chyzy00,chyzy03,gaensler05,kepley07}).
%% where the effects of rotation are
%% negligible like the solar tachocline\cite{}
The case of Keplerian rotation 
corresponds to the local shearing-sheet model of thin accretion disks,
for which the angular velocity decreases radially $\propto R^{-3/2}$.
It is commonly used to study turbulence and transport in accretion disks,
in particular the MHD turbulence driven by the magneto-rotational instability (MRI) 
\citep[see, e.g.,][and references therein]{balbus03,fromang07,lesur07}.

\runtable

We solve \eqsand{eq:ns}{eq:induction} in a shearing periodic computational domain
by means of a Lagrangian spectral method \citep{ogilvie98,lithwick07}. 
In our model, we use a random body force $\vf$ that is
\emph{nonhelical} satisfying $\vf\cdot\(\vdel\times\vf\)=0,$
$\delta$-correlated in time (white noise), 
and has a characteristic length scale $l_f=2\pi/k_f$. 
Choosing $\vf$ to be $\delta$-correlated in time makes it possible to 
impose the average injected power $\eps=\tavg{\<\vu\cdot\vf\>}$ precisely. 
Here and in what follows, 
angular brackets denote volume averaging, while overbars represent time
averaging. The characteristic outer scale length of the
turbulence excited by our random body force is $\sim l_f$ and for a given viscosity, 
the corresponding outer scale velocity $\sim\urms=\<u^2\>^{1/2}$ 
is then determined by $\eps$; 
for sufficiently small viscosities, dimensionally, we must have 
$\urms\sim(\eps l_f)^{1/3}$. 

In astrophysical systems, where the source of the turbulence is often related to 
the large-scale shear itself, the most relevant case tends to be $S\tau\sim 1$, 
where $\tau\sim l_f/\urms$ is the turnover time of the turbulent motions 
at the outer scale. In our model, we have control over the turbulence via 
the parameters of the body force described above. 
This allows us to consider a range of values of $S$. 
Such a study is useful because $S\tau\sim1$ is only an order-of-magnitude 
relation (so it is not obvious which precise value of $S$ would be 
the most ``realistic'') and also because any analytic theory of the 
shear dynamo would have to predict how the effect depends on the 
value of $S$ (usually it is only the asymptotic case 
$S\tau\ll1$ that can be treated analytically, as, e.g., in 
\citealt{RK03,RK04}).

As we shall see, a significant separation
in scales between $l_f$ and the dimensions of the computational domain
is necessary to get magnetic field growth---this is particularly true 
for weak shears. Also very long run times (at least hundreds and, for weak shears, 
thousands of turnover times) are needed to get reliable statistics. 
This is computationally expensive and we were only able to satisfy both
requirements within available computational resources by restricting our study
to computational domains elongated in the $z$ direction. 
Thus, in all the runs presented here, we have $L_x=L_y=1$, 
the forcing scale is $l_f=1/3$ and $L_z$ is taken to be between 
$8$ and $128$ (depending on the value of $S$), so $L_z\gg l_f$.
The minimum value of $L_z$ needed for a given simulation to produce
results independent of the box size in the $z$ direction depends on
the shearing rate (this issue is discussed in \secref{sec:convergence}).  
Setting $L_x=L_y=1$ together with $\eps=1$ defines the code units of 
length and time. 

A list of all runs is given in \tabref{tb:runs}.
The S (sheared, nonrotating) and K (Keplerian) runs have $\nu=\eta=10^{-2}$. 
The S runs have $\urms=1.0\pm0.3$ while
the K runs have $\urms=.5\pm 0.06$ 
(larger velocities in the former case are due to the excitation of large-scale velocity structures 
in the absence of rotation; see \secref{sec:nlinst}). 
This corresponds to Reynolds numbers $\Re=\urms/k_f\nu\sim5$ and $3$ for 
the S and K runs, respectively. The magnetic Reynolds number 
$\Rm=\urms/k_f\eta=\Re$ for all runs, as the study of the dependence 
of the shear dynamo properties on the magnetic Prandtl number 
$\Pm=\Rm/\Re$ lies outside the scope of this paper. 
As the Reynolds numbers are low, only relatively low resolution is 
needed and we find it sufficient to have 32 collocation points 
per code unit of length. Strictly speaking, this is not 
fully developed turbulence as there is little, if any, 
separation between the energy-containing (outer) and viscous scales. 
However, this is not a serious limitation because 
one expects the mean field dynamo mechanism and 
the properties of the resulting mean field 
to depend only on the outer scale of the turbulence 
(rather than on the structure of the inertial-range) 
and, therefore, to become asymptotically independent
of $\Re$ for some critical $\Re \gtrsim 1$ 
While one cannot, of course, always take this kind of intuition for granted, 
one expects it to work at least 
for a turbulence driven with fixed input power $\eps$ as in our simulations. 
It does, indeed, appear to work: 
by rerunning a selection of the runs with lower viscosity, 
we have confirmed that the results reported below are independent 
of $\Re$. 

All simulations start with a dynamically insignificant random 
seed magnetic field 
whose volume average is (and remains) zero. 
The initial magnetic energy is 
$\<B^2\>$$\sim$$10^{-20}\,\urms^2$ for runs S1--S13 and 
$\<B^2\>$$\sim$$10^{-24}\,\urms^2$ for runs K1--K8 and FD1--FD2.  
The shear dynamo mechanism generates a magnetic field with large-scale 
structure in the $z$ direction. The field grows exponentially until it
reaches a saturated state, in which the value of $\<B^2\>$ becomes comparable 
to $\<u^2\>$. In this paper, we only analyse the kinematic regime of this
dynamo, in which the back reaction of the magnetic field onto the
velocity field through the Lorentz force can be neglected, and so only
consider data for times before $\<B^2\>$ reaches $10^{-4}\<u^2\>$.  
All numbers reported in \tabref{tb:runs} are for this kinematic stage, 
except for the run FD1 (see \secref{sec:fd}). 

\begin{figure}
{\includegraphics[width=.49\textwidth]{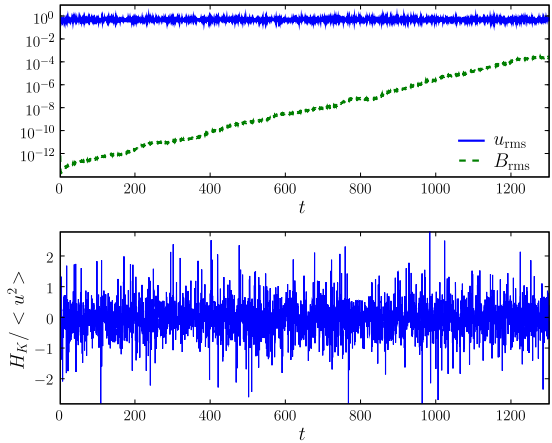}}
\caption{(online colour at: www.an-journal.org) The upper panel shows the evolution of $\Brms$ and $\urms$ 
with time for the run K2 ($S=-1, \Omega=2/3$). The lower panel shows 
the normalised kinetic helicity as a function of time $t$ for the same run.
Fluctuations in kinetic helicity only occur on turbulent time scales and have 
zero time average. This is true for runs both with and without rotation.}
\label{fig:helicity}
\end{figure}

Before we close this section, several important remarks 
are in order, concerning the physical effects that are 
{\em not} present in our simulations. 

First, the magnetic Reynolds numbers $\Rm$ for all
runs except FD1 and FD2 are deliberately chosen to lie 
\emph{below} the critical threshold 
value $\Rmc$ at which the fluctuation dynamo sets in. 
This ensures that the magnetic-field growth observed
in the S and K runs is a result of a ``pure'' mean-field dynamo, 
unpolluted by the fluctuation dynamo and that the operation of 
the former is not predicated on the existence of the latter. 

Second, the magnetorotational instability, MRI turbulence or MRI dynamo are not
present in the Keplerian runs K and FD1.  This is because the magnetic field is
extremely weak (in the K runs) and the viscosity is large.  In addition, we
know that the MRI dynamo and turbulence have so far only been found for values
of $\Pm$ systematically larger than unity, in high-resolution simulations of
Keplerian shearing boxes elongated in the $y$ direction
\citep{fromang07,lesur08}. 

Third, the turbulence in all our simulations is statistically 
nonhelical, even for the runs with rotation. 
The kinetic helicity $H_K=\<\vu\cdot\(\curl\vu\)\>$ 
has a zero time average, $\overline{H_K}=0$, 
and only fluctuates on the turbulent time 
scales ($\sim\tau\sim l_f/\urms$). 
This is a consequence of $\vf$ being nonhelical, ${\vf\cdot\(\curl\vf\)}=0$,
and $\delta$-correlated in time. 
The latter guarantees that the time average of helicity production is
$\tavg{\(\curl\vu\)\cdot\vf}+$ $\tavg{\vu\cdot\(\curl\vf\)}=0$.  
\Figref{fig:helicity} shows the time evolution of $H_K$ for a
typical run. This excludes the possibility that $\alpha$ effect, 
at least in its standard and simplest form, 
is present in our simulations.

\section{Shear dynamo without rotation}
\label{sec:linear}

In this section, we document the numerical results obtained for $S\neq 0$
and $\Omega=0$ (turbulence $+$ shear, no rotation). This regime is 
covered by the S runs (\tabref{tb:runs}). In all of them, we find 
that magnetic field grows exponentially. 
Let us first describe the spatial structure of this growing field, 
then the dependence of its properties on the shearing rate $S$
and their convergence with respect to the vertical size of 
the computational box. 

\subsection{Structure and evolution of the growing field}

\begin{figure*}
\begin{flushright}
  \includegraphics[width=.95\textwidth]{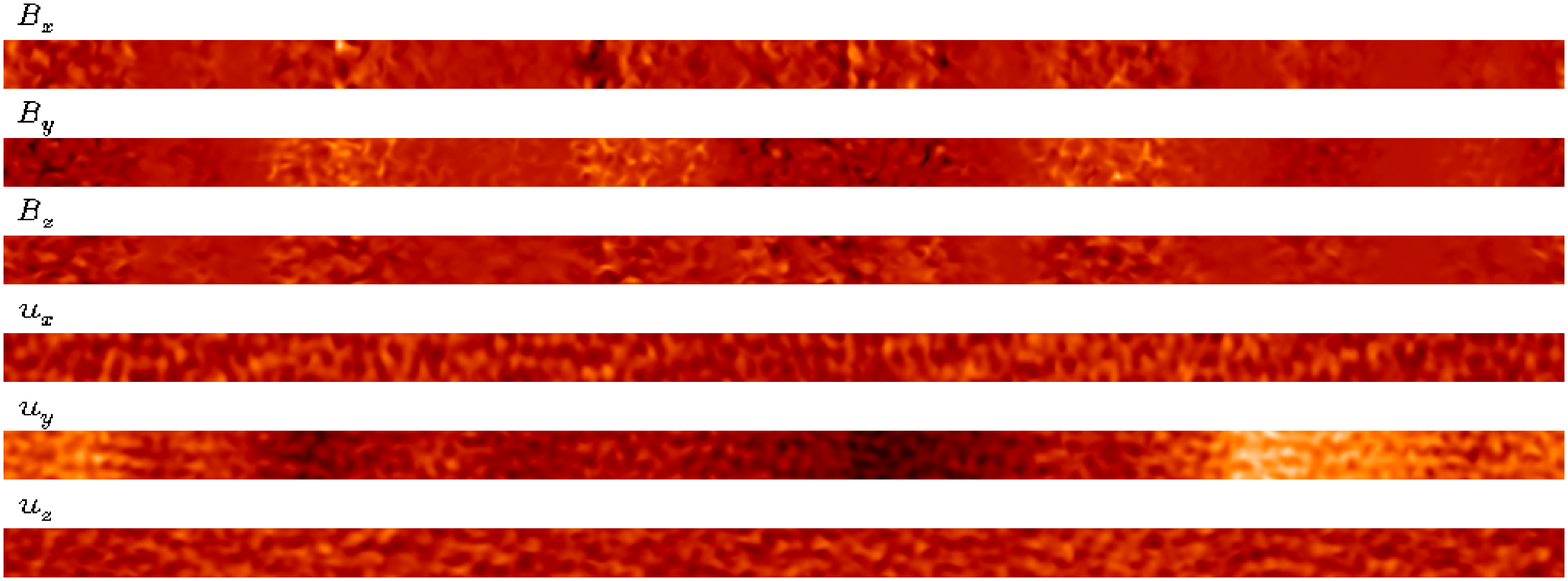} 
\end{flushright}

\includegraphics[width=1.05\textwidth]{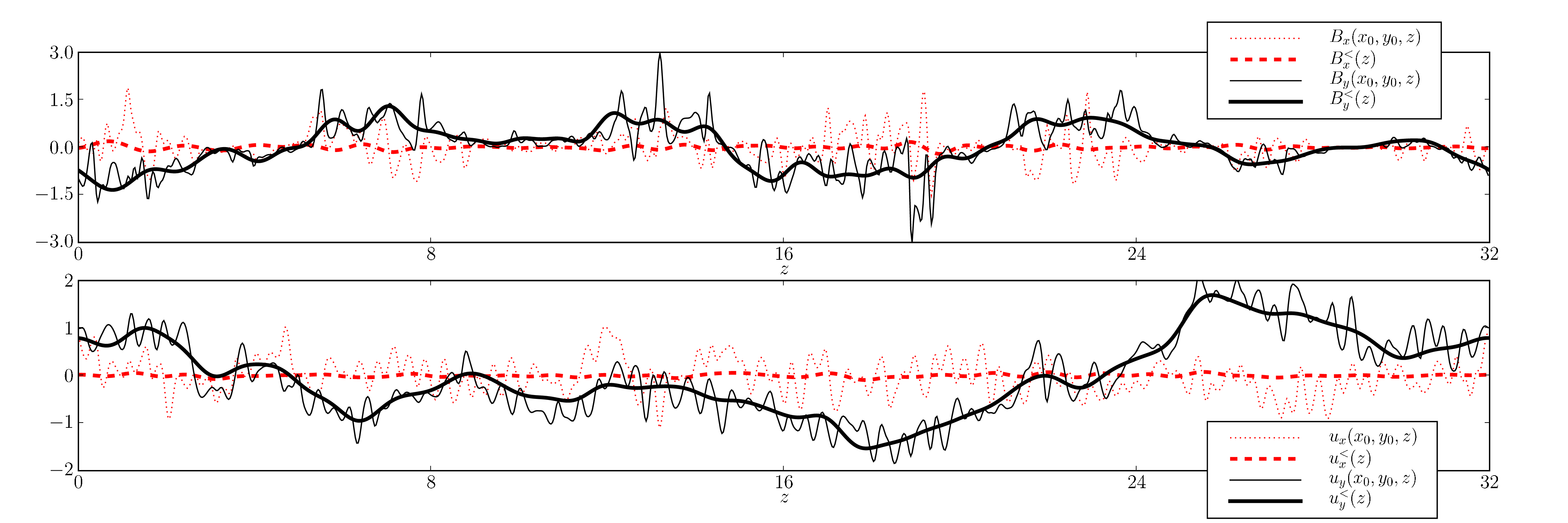} 
\caption{(online colour at: www.an-journal.org) Snapshots of $B_x$, $B_y$, $B_z$, $u_x$, $u_y$ and $u_z$ in the 
$(y, z)$ plane for run S5 ($S=-1$, $\Omega=0$) at $t=300$.  
Both velocity and magnetic fields have structure in the $z$ direction 
with characteristic scale far larger than the horizontal box size 
$L_x=L_y=1$, in addition to the turbulent fluctuations that have 
characteristic length scales smaller than $1$
(see \secref{sec:nlinst} regarding the velocity structure). 
The panels underneath the snapshots show the instantaneous 
profiles of the field components (for a given
$x=x_0$ and $y=y_0$) vs.~$z$ (thin lines) and their
large-scales means Fourier-filtered 
according to \eqref{Bm_def} (thick lines). 
}
\label{fig:structure}
\end{figure*}

\Figref{fig:structure} shows a cross section of the different components of
$\vB$ and $\vu$ from a representative run with no rotation and weak shear.
One can clearly distinguish between fluctuations with scales comparable to the
forcing scale (which is, on the average, $l_f=1/3$ of 
the horizontal box size $L_x=L_y$) and a large-scale ``mean'' field.  
The small-scale fluctuations cannot result from the fluctuation 
dynamo because $\Rm<\Rmc$. 
Instead, they stem from the tangling of the mean field by the
turbulence -- the energy in these fluctuations 
is comparable to the energy of the mean field 
\citep[for larger $\Rm$, this energy will be even larger; 
see, e.g.,][and references therein]{sch07}. 

\begin{figure}
\includegraphics[width=.49\textwidth]{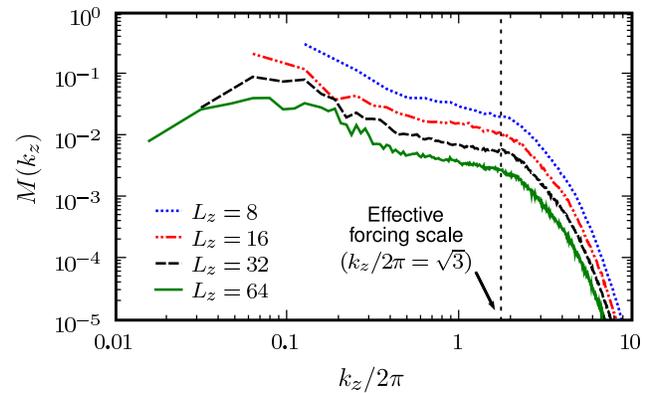} 
\caption{(online colour at: www.an-journal.org) Time averaged normalised one-dimensional spectra of the magnetic 
energy (Eq. 3) for runs S3--S6 ($S=1$, $\Omega=0$, vertical box 
sizes $L_z=8,16,32,64$). As $L_z$ is increased, the peak of the spectrum 
settles at a wave number independent of $L_z$ (see \secref{sec:convergence}).}
\label{fig:spectrum}
\end{figure}

The spatial structure 
of the magnetic field is illustrated by \figref{fig:spectrum}, which 
shows the time-averaged normalised spectrum of the growing field with 
respect to $k_z$, 
\be
\label{M_def}
M(k_z) = \overline{{1\over\<B^2\>}\sum_{k_x,k_y}{|\vB(k_x,k_y,k_z)|^2}}.
\ee 
Note that since the average forcing wave number is 
$k_f=(k_{x,f}^2+k_{y,f}^2+k_{z,f}^2)^{1/2}=2\pi/l_f$, where $l_f=1/3$, 
the mean square forcing wave number associated with one 
spatial dimension is $k_{z,f}=(k_f^2/3)^{1/2}=2\pi\sqrt{3}$. 
The peak of the spectrum at $k_z\ll k_{z,f}$ corresponds to the mean field. 
The fluctuations at smaller scales are also evident, although at the 
low value of $\Rm$ that we use they are not very large. 

In order to systematically 
isolate the mean field from the small-scale fluctuations, 
we filter out the highest Fourier modes:
\be
\label{Bm_def}
\vBm(z) = \sum_{k_z/2\pi < 1} \vB(k_x=0,k_y=0,k_z)\, {\rm e}^{ik_zz} .
\ee
This is our definition of the mean field. 
The Fourier filtering averages out all variation in the $x$ and $y$ directions 
and removes the scales in the $z$ direction smaller than $L_x=L_y=1=3l_f$. 
Note that $\Bm_z$ is strictly zero because $\vdel\cdot\vB=0$: indeed,
$\dd_z\Bm_z=-\dd_x\Bm_x-\dd_y\Bm_y=0$ as the filtered field has no variation in
$x$ and $y$.  Of the two nonzero components of the growing mean field, $\Bm_y$
component is the larger: this can be seen in \figref{fig:structure}, where we
plot the instantaneous profiles of the field components and their
Fourier-filtered means.   
This relative predominance of $B_y$ is not surprising given 
the form of \eqref{eq:induction} where in addition 
to the turbulent stretching terms, there is a linear term $B_xS\vy$ whereby the 
shear systematically produces $B_y$ from $B_x$ (azimuthal field from radial 
field in axisymmetric systems; this is known as the $\Omega$-effect 
in dynamo theory). 

\begin{figure*}[t]
\includegraphics[width=.49\textwidth]{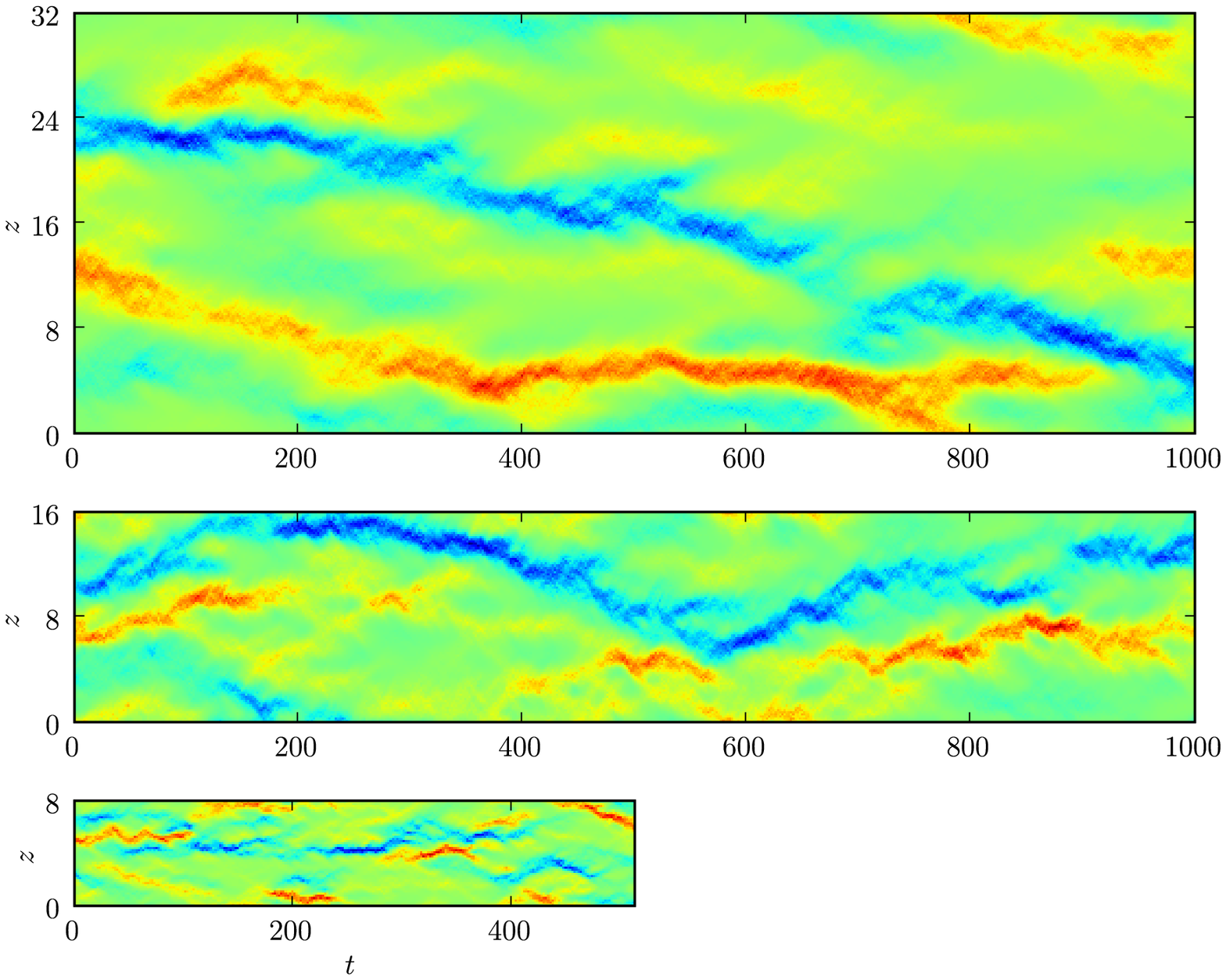}\quad 
\includegraphics[width=.49\textwidth]{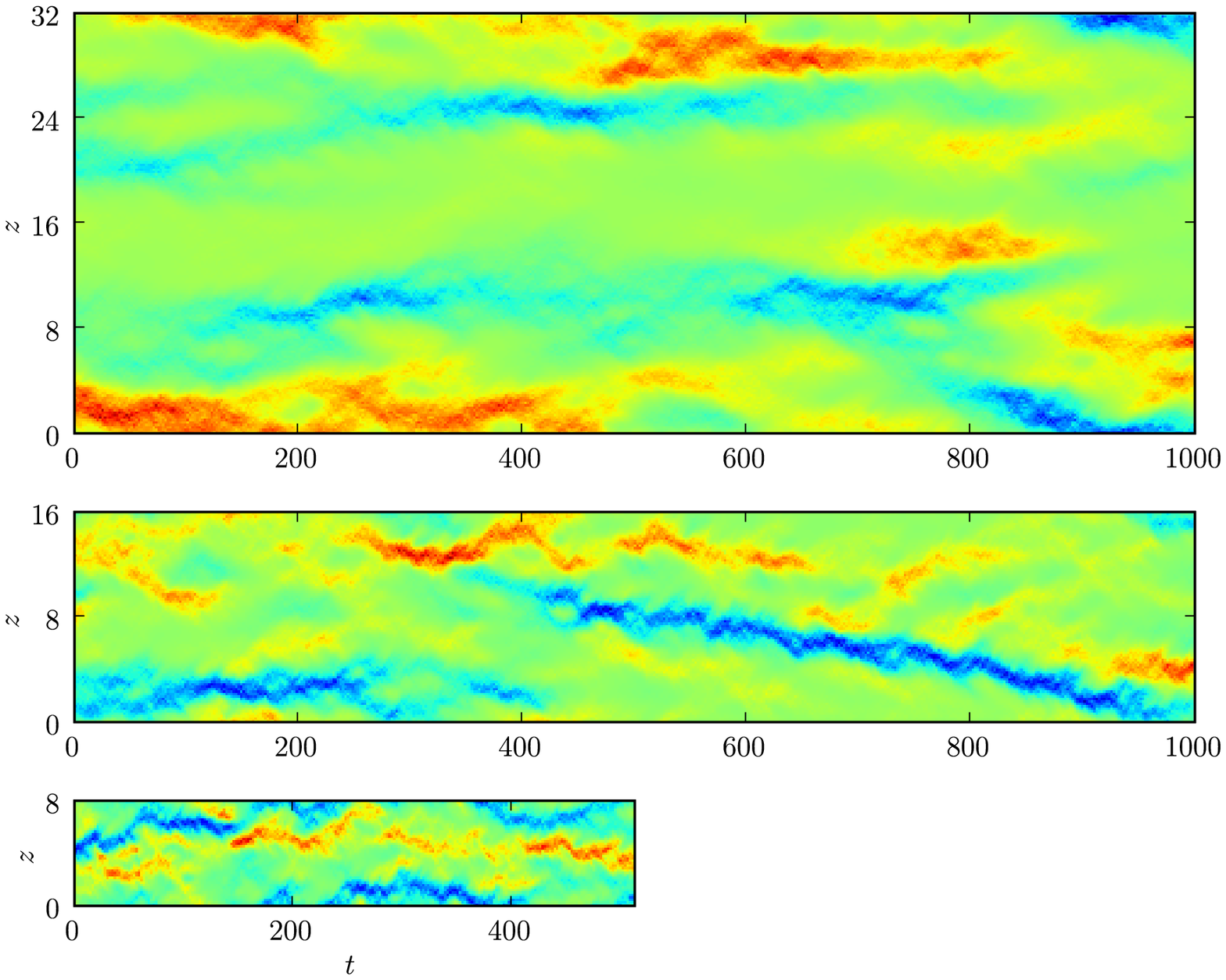}
\caption{(online colour at: www.an-journal.org) \emph{Left panel}: the $y$ component of the magnetic field averaged over $x$
and $y$,  $\xyavg{B_y}(z,t)$,  for three runs with linear shear and $S=1/2,1,2$
and $\Omega=0$ (runs S8, S4 and S1, 
from top to bottom). 
The characteristic length scale $\lb$ (manifested by variations along the
vertical axis of the figures) decreases with increasing shearing 
rate $S$. The magnetic structures are correlated over times 
(variations along the horizontal axis of the  figures) that are very
long compared to the turnover time of the turbulence $\tau\sim 1/3$. 
\emph{Right panel}: similar figures for the case with Keplerian rotation 
(runs K1, K2, K3) discussed in \secref{sec:kepler}.}
\label{fig:scales}
\end{figure*}

As well as having a large-scale spatial coherence, the mean 
field is also coherent over long times. While it does
not seem to be an eigenmode in the usual sense of having 
a shape that is exactly constant in time, the characteristic times over 
which it changes are much 
longer than both the turbulent turnover time $\tau$ and the shearing time $S^{-1}$. 
The time evolution of the mean field is illustrated by \figref{fig:scales} 
(the left panels for nonrotating runs, the right panels 
for rotating runs; see \secref{sec:kepler}). Here we plot the evolution of 
the  
profile of the $y$ component of the magnetic field averaged 
over $x$ and $y$, $\xyavg{B_y}(z,t)$.
In all cases, we see a well-defined evolving spatial structure: patches of 
stronger or weaker field slowly drift along $z$, occasionally peter out 
and reemerge. 

\begin{figure*}[t]
\includegraphics[width=.49\textwidth]{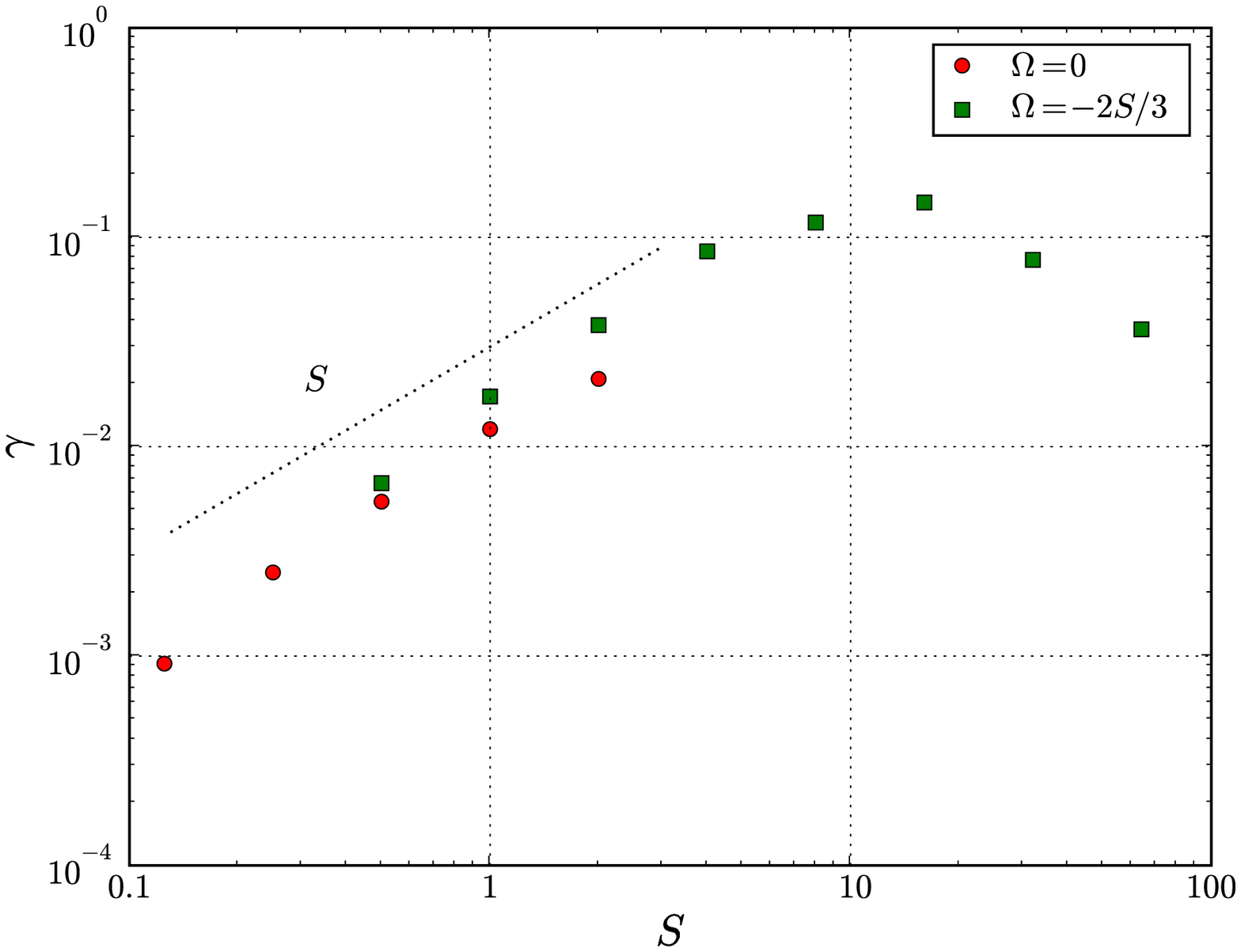}\quad
\includegraphics[width=.49\textwidth]{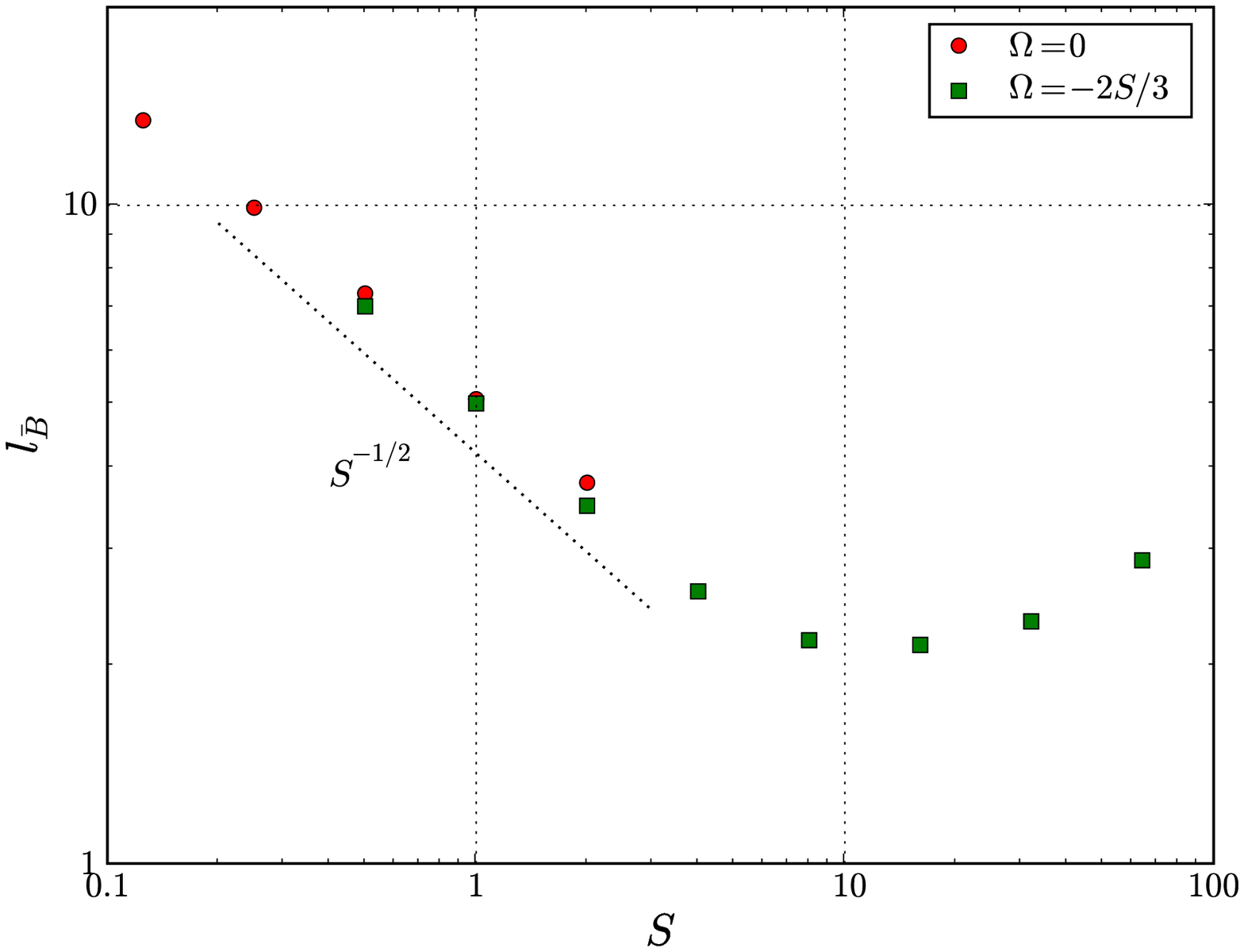}
\caption{(online colour at: www.an-journal.org) \emph{Left panel}: the growth rate $\gamma$ of $\Brms$ vs.\ the 
shearing rate $S$. The numerical values of $\gamma$ are given in \tabref{tb:runs}. 
For each value of $S$, only the
run with the largest value of $L_z$ is shown in the figure to avoid clutter.
The plotted values correspond to 
the nonrotating runs S2, S6, S9, S11, S13 (red circles; see \secref{sec:shearscan}) 
and the Keplerian runs K1--K8 (green squares; see \secref{sec:kepler}). 
\emph{Right panel}: the characteristic scale $\lb$ of the mean field, as defined 
by \eqref{lb_def} vs.\ the shearing rate for the same runs
(see \tabref{tb:runs} for the numerical values of $\lb$). 
Note that only small shear values are accessible in the
nonrotating case because larger shears encounter nonlinear
instability in that regime (see \secref{sec:nlinst}). 
In the Keplerian regime, these nonlinear
instabilities are suppressed, which makes it possible to explore
regimes with larger shear (see \secref{sec:kepler}).}
\label{fig:scaling}
\end{figure*}

\subsection{Dependence on the shearing rate}
\label{sec:shearscan}

Dependence of the properties of the growing field on the shearing rate 
$S$ is the first obvious parameter scan to carry out. It provides 
a quantitative test of past mean-field theories and possibly 
a valuable hint as to how one might construct future ones. 

The dependence of the growth rate $\gamma$ of $\Brms = \<B^2\>^{1/2}$ 
on $S$ is plotted in the left panel of \figref{fig:scaling}. 
The growth rates for the nonrotating runs appear to fit 
approximately the dependence $\gamma\propto S$ for $S$ between $1/8$ and $2$. 
Note, however, that the actual numerical values of $\gamma$ are 
substantially smaller than $S$ (see \tabref{tb:runs}). 
We do not currently know what determines 
the proportionality coefficient (we have checked that it is not the value of
$\Re$: the growth rates presented here do not change as 
the Reynolds number is increased). 

In \tabref{tb:runs}, we also list $\tavg{[\int {\Bm_y}^2 \diff z/\int {\Bm_x}^2
\diff z ]^{\frac12}}$.  
Its value is approximately 10 irrespective of  $S$ , although it shows a
slight tendency to increase with $S$. 
This is consistent with $\gamma\propto S$ because from the $y$ component of the
induction equation \eqref{eq:induction}, we can estimate
$\dd_t\Bm_y\sim\gamma\Bm_y\sim S\Bm_x$, so ${\Bm_y/\Bm_x}\sim
{S/\gamma}\sim\const$.

From \figref{fig:scales}, we already knew that 
the scale of the mean field decreases with shear.
In order to study this dependence quantitatively,
we define the characteristic length scale of the mean field 
as follows 
\be
\label{lb_def}
\frac 1\lb= \tavg{ \left[ \frac {\int \! \diff z (\partial  
\Bm_y/ \partial z)^2} {\int \! \diff z {\Bm_y}^{2}} 
 \right]^{1/2} }.
\ee
The values of $\lb$ for all runs are
listed in \tabref{tb:runs} and plotted in the right panel of
\figref{fig:scaling} as a function of the shearing rate $S$. 
We see that for the values of $S$ that we have studied, 
the characteristic length scale $\lb$ approximately matches 
the scaling $1/\sqrt{S}$. It is possible to argue 
\citep{yousef08} that for a generic model form of mean-field equations, 
this scaling would indeed be consistent with the linear scaling 
of the growth rate, $\gamma\propto S$. 

These scalings do not agree with the mean field theory in its current state. 
Namely, \citet{RK03}, that proposed the existence of the 
shear dynamo using a mean-field theory with 
the $\tau$-approximation closure to incorporate the effect 
of turbulence, predicted that for the fastest-growing mode  
$\gamma\propto S^2$ and $\lb\propto 1/S$ for $S\tau\ll 1$. 
These scalings do not seem to match the numerical results. 
However, we cannot exclude the possibility that the values of $S$ considered 
by us are, in fact not asymptotic in the small parameter $S\tau$ 
and that Rogachevskii and Kleeorin's scalings might hold for even weaker 
values of shear or for numerical domains that allow more scale separation 
between the turbulence and the horizontal box size ($L_x$ and/or $L_y$). 

\subsection{Independence of the computational domain size}
\label{sec:convergence}

\begin{figure}[t]
\includegraphics[width=.49\textwidth]{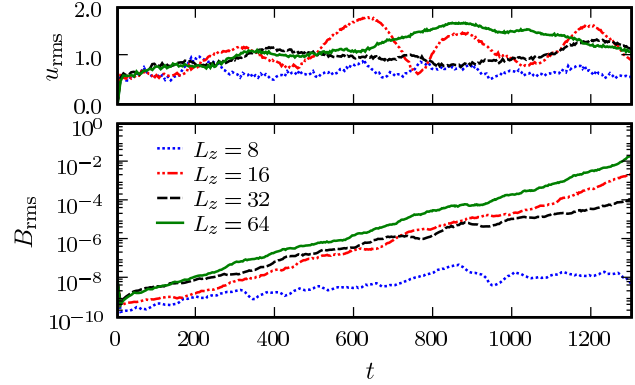} 
\caption{(online colour at: www.an-journal.org) \emph{Bottom panel}: time evolution of $\Brms$ for a set of
runs with $S=1$ and varying vertical box size $L_z=8,16,32,64$ 
(runs S3--S6). When $L_z$ is sufficiently large (here 16) 
the growth rate $\gamma$ becomes independent of $L_z$. 
\emph{Top panel}: evolution of $\urms$ for the same runs. 
The velocity fields develops large long-lived fluctuations (see \secref{sec:nlinst}).} 
\label{fig:s1-evo}
\end{figure}

It is important to ensure that the results reported above, particularly
those of \figref{fig:scaling}, are independent of the size of the
computational domain. Since the growing mode has large-scale structure
in the $z$ direction, we do this by performing, for each value of $S$,
a set of runs in boxes with successively larger values of $L_z$  until
we can ascertain that the properties of the growing mode are
independent of $L_z$. To illustrate this procedure, the bottom panel of
\figref{fig:s1-evo} shows the time evolution of $\Brms$ for runs with $S=1$: 
exponential growth with a growth rate independent of $L_z$ emerges as 
$L_z$ is increased. Another illustration of the convergence with 
respect to $L_z$ is \figref{fig:spectrum}, where normalised magnetic-energy spectra 
are plotted for the same runs as in \figref{fig:s1-evo}. We see that 
the peak of the spectrum (corresponding to the mean field) becomes 
independent of $L_z$ at sufficiently large $L_z$. 

We have performed this type of study for a number of values of $S$ 
and in all cases confirmed convergence with respect to $L_z$. 
This is documented in \tabref{tb:runs}. 
Naturally, as the scale of the mean field decreases with increasing shearing 
rate $S$ (\secref{sec:shearscan}), the minimum $L_z$ required for convergence 
also decreases. Thus, we find that $L_z=8,16,32,64$ 
are sufficiently large for shearing rates $S=2,1,1/2,1/4$ and $1/8$, 
respectively. 

The convergence with respect to $L_z$ ensures that the properties of the mean
field and their dependence on the shearing rate $S$  are physical and are not
numerical artifacts related to the choice of the computational domain size in
$z$. 
It is also important to check whether the results depend on the dimensions of
the computational domain in the $x$ and $y$ direction. Such study, which is
more computationally  demanding, has been left outside the scope of this paper
and will be undertaken elsewhere.

\subsection{Vorticity dynamo} 
\label{sec:nlinst}

\begin{figure}[t]
\includegraphics[width=.49\textwidth]{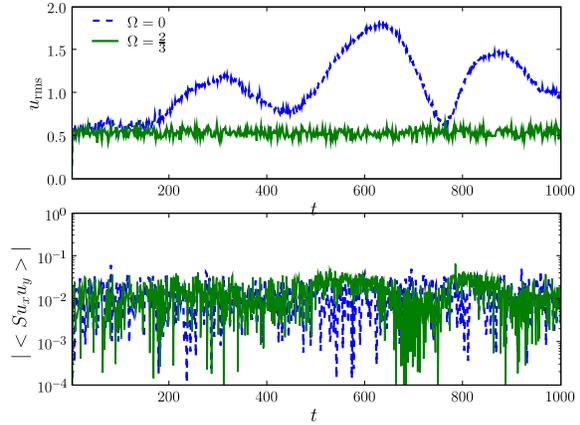} 
\caption{(online colour at: www.an-journal.org) \emph{Top panel}: time evolution of $\urms$ for the 
runs S4 ($S=-1$, $\Omega=0$; dashed line) and K2 ($S=-1$, $\Omega=2/3$; solid
line).  
Large-amplitude long-lived in $\urms$ can be seen for run S4, but not for run
K2.  \emph{Bottom panel}: evolution of the power injected into the flow by shear
$-S\<u_yu_x\>$ for the same runs. This is to be contrasted 
with the power injected by the forcing, $\eps=1$. 
} 
\label{fig:bumps} 
\end{figure}

We finally mention another noteworthy result in the nonrotating
regime. We observed that the velocity field  also develops energetic
fluctuations with correlation times much greater than $\tau$ and
$S^{-1}$. This property is illustrated in the top panel of 
\figref{fig:s1-evo}. Since homogeneous linear shear is known to become nonlinearly
unstable \citep[see][and references therein]{pumir96} at large enough $\Re$, 
one might imagine that the observed velocity fluctuations result from 
finite-amplitude destabilisation of shear by forcing-scale motions. 
Indeed, for high enough values of $S$ and/or $L_z$ we experienced that our
background shear became vigorously unstable and developed into a flow
characterised by $\urms \gg 1$ and  fluctuations on all
scales. However, this is not the case for any of the runs presented here. 
There is a key difference between these runs and the truly 
unstable cases. Finite-amplitude
instabilities are known to feed on shear \citep{casciola03} 
and are characterised by large $\<u_xu_y\>$ correlations associated 
with turbulent momentum transport. 
This contrasts with our results: as \figref{fig:bumps}
shows, there seems to be almost no correlation between the large velocity 
fluctuations and the mean power injected into the flow by the shear
$-S\<u_xu_y\>$ in our runs. The ratio 
of this shear power to the forcing power $\eps$ stays
much smaller than unity, $-S\<u_xu_y\>/\eps\ll1$, 
implying that the velocity fluctuations feed on the forcing
and not on the shear. 

These observations suggest 
that the finite-amplitude instabilities associated with the shear are not the cause
of the observed velocity fluctuations. Given the
formal similarities between the induction and vorticity equations,
one might argue that a ``vorticity dynamo'' similar to the shear dynamo 
is triggered \citep{EKR03}. This is supported by the fact that the 
large-scale velocity structure forms mainly in $u_y$ (see \figref{fig:structure}), 
which corresponds to the large-scale vorticity $\omega_x = -\dd_z u_y$, the quantity one 
expects to be analogous to $B_y$. 
The difference between the vorticity and the magnetic-field dynamos 
is that the ``seed'' field for the vorticity dynamo results from the forcing
and is, therefore, never dynamically insignificant. Thus it is not
possible to observe a long kinematic growth phase for this vorticity
dynamo. One might conjecture that the large velocity fluctuations 
represent the nonlinear regime of such a dynamo. 
We note, however, that the status of the vorticity dynamo remains 
uncertain because, as seen, e.g., in \figref{fig:s1-evo}, both 
the time duration of the large velocity fluctuations and their 
amplitude depend strongly on $L_z$. Thus, unlike for the 
magnetic field, there is no convergence here with respect to the 
domain size and we must remain cautious in interpreting these results. 

Finally, we observe that the emergence of 
the large-amplitude, long-lived, large-scale fluctuations in the 
velocity field does not appear to be strongly correlated 
with the operation of the shear dynamo: compare, e.g., the time 
evolution of $\urms$ and $\Brms$ shown in \figref{fig:s1-evo}. 
As we are about to see, the absence of these velocity fluctuations 
in the simulations with Keplerian rotation (see \figref{fig:bumps}) 
does not change any of the 
basic properties of the shear dynamo, so we feel safe in ruling out 
the possibility that the large-scale velocity fluctuations are 
a required ingredient in the shear dynamo. 

\section{Shear dynamo in rotating systems}
\label{sec:kepler}

Since many astrophysical plasmas (stars, disks, galaxies) are differentially rotating, 
it is important to understand whether, and how, the shear dynamo is affected by 
imposing a uniform rotation of the system.
We focus on the case of Keplerian rotation introduced in \secref{sec:model},
because it is relevant for accretion-disk dynamos and 
has been extensively studied numerically in this context 
\citep[e.g.,][and references therein]{balbus03,fromang07}. 
This regime is thought to be nonlinearly stable from the hydrodynamic point of
view \citep{lesur05,ji06,rincon07}, meaning 
that the nonlinear instabilities characteristic of nonrotating shear flows 
are absent. They are, indeed, absent in our simulations.  
Furthermore, as illustrated by \figref{fig:bumps}, even 
the large velocity fluctuations present in the time series of $\urms$ 
for the nonrotating runs are suppressed in the case of Keplerian rotation. 
The suppression of these fluctuations, which were tentatively interpreted in terms 
of a ``vorticity dynamo'' in \secref{sec:nlinst}, is likely to be another consequence 
of the stabilising effect of Keplerian rotation on hydrodynamic motions. 

As we have mentioned in \secref{sec:nlinst}, the 
nonlinear destabilisation of the shear for too large values 
of $S$ and $L_z$ imposed limitations on the range of values of 
$S$ for which shear dynamo could be studied in the nonrotating case. 
The absence of such destabilisation in the simulations with 
Keplerian rotation allows us to extend the study the shear dynamo to 
much higher values of $S$ than in the nonrotating case (see the K runs in 
\tabref{tb:runs}). 

The main conclusion of this study is that Keplerian rotation does not seem 
to alter the properties of the shear dynamo in any significant way. 
All the results concerning the growing magnetic field
reported in \secref{sec:linear}, both qualitative and quantitative,  
continue to hold in the rotating case. 
The right panel of \figref{fig:scales} shows the evolution 
of the large-scale field with time for three runs with rotation.  
It is hard to see any qualitative difference between the 
structure and evolution of the growing field in the rotating case and 
its structure in the nonrotating case shown in the left panel of the
same figure. The similarity between the two cases is further demonstrated in
\figref{fig:scaling}, which shows that for low values of $S$, both
the growth rates $\gamma$ and the characteristic length scales $\lb$ 
seem to follow the same scaling laws in the rotating and nonrotating cases. 
The numerical values of $\gamma$ with and without rotation are quite close, 
although the growth rates in the presence of rotation appear to be 
systematically slightly higher. This small difference may result from the 
changes in the structure of the turbulence caused by rotation or it could 
be due to an additional contribution to the generation of magnetic field 
from other mean-field dynamo mechanisms associated with rotation: 
e.g., the R\"adler, or $\bmath{\Omega}\times\bmath{J}$, effect 
\citep{raedler69,raedler03}. It is worth reemphasising in this context 
that the turbulent motions driven by our
forcing remain nonhelical on average even in the presence of Keplerian
rotation (see \figref{fig:helicity} and the discussion at end of 
\secref{sec:model}), so that the rotating shear dynamo does not 
involve an $\alpha$ effect in the usual sense. 

\begin{figure}
\includegraphics[width=.49\textwidth]{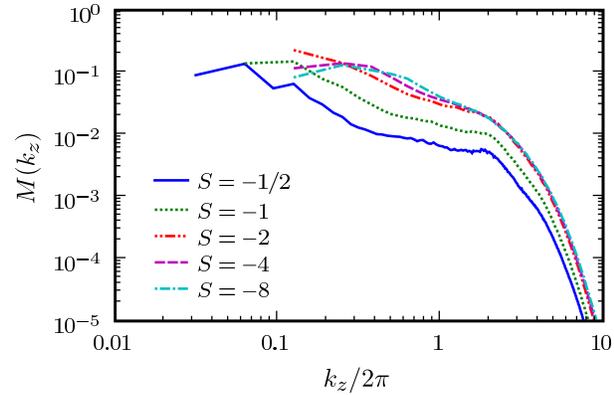} 
\caption{(online colour at: www.an-journal.org) Time averaged normalised one-dimensional spectra of 
magnetic energy (Eq. 3) for runs K1--K5 ($S=-\frac12$, $-1$, $-2$, $-4$, $-8$;
 $\Omega=-2S/3$).
As $S$  decreases, the peak of the spectrum, corresponding to the mean field,
moves towards lower wave numbers, corresponding to larger scales.}
\label{fig:spectrarot}
\end{figure}

\Figref{fig:scaling} shows that, as $S$ is increased beyond the values studied
for $\Omega=0$, the scalings of the growth rate and the mean-field scale,
$\gamma\propto S$ and $\lb\propto1/\sqrt{S}$, continue until 
$\gamma$ reaches its maximum and $\lb$ its minimum at $S\sim10$. 
At even larger shears, the growth rate starts to decrease, as it 
intuitively should. Indeed, the dynamo must surely disappear 
in the limit of infinite shear. More precisely, in the limit when 
the shearing rate associated with the imposed shear is much larger 
than the turbulent rate of strain, $S\gg|\vdel\vu|$, the nonlinear terms 
(the terms involving turbulent velocities) in the 
induction equation \eqref{eq:induction} become negligible---without 
these nonlinearities, we cannot have an exponentially growing 
magnetic field. 

This argument suggests that the qualitative change 
associated with the peak of the growth rate in \figref{fig:scaling} 
corresponds to the transition from the weak-shear 
regime ($S\tau\ll1$) to the strong-shear regime ($S\tau\gg1$). 
Quantitatively, this transition evidently occurs at $S\sim10$. 
Once the imposed shear $S$ becomes stronger than the 
turbulent rate of strain at the forcing scale $\tau^{-1}\sim\urms/l_f$, 
smaller turbulent scales should come into play. 
In order to understand what happens 
when the shear is strong, one must analyse the changes that 
such a shear causes to the inertial-range turbulence, 
determine the transition scale at which the turbulent rate of strain 
catches up with the shear, etc. Since the numerical simulations presented here, 
unlike real turbulent systems, have no extended inertial range, their relevance 
in the strong-shear regime is questionable and we shall not discuss 
this regime here. 

Finally, for completeness, in \figref{fig:spectrarot} we show the one-dimensional 
magnetic-energy spectra, defined by \eqref{M_def}, for Keplerian runs 
with several values of shear. We see that, as $S$ is decreased, 
the peak of the spectrum, corresponding to the mean field, moves 
towards lower wave numbers (corresponding to larger scales; 
cf.\ \figref{fig:scaling}). The rest of the spectrum represents 
magnetic fluctuations produced by the turbulent tangling 
of the mean field. Here, like for the nonrotating runs 
in \secref{sec:linear}, the tangling, or magnetic induction
\citep[see][and references therein]{sch07}, is the only mechanism 
for producing small-scale fields because $\Rm$ is chosen below the threshold 
for the fluctuation dynamo. 

\section{Shear dynamo and fluctuation dynamo}
\label{sec:fd}

\begin{figure*}
\includegraphics[width=.49\textwidth]{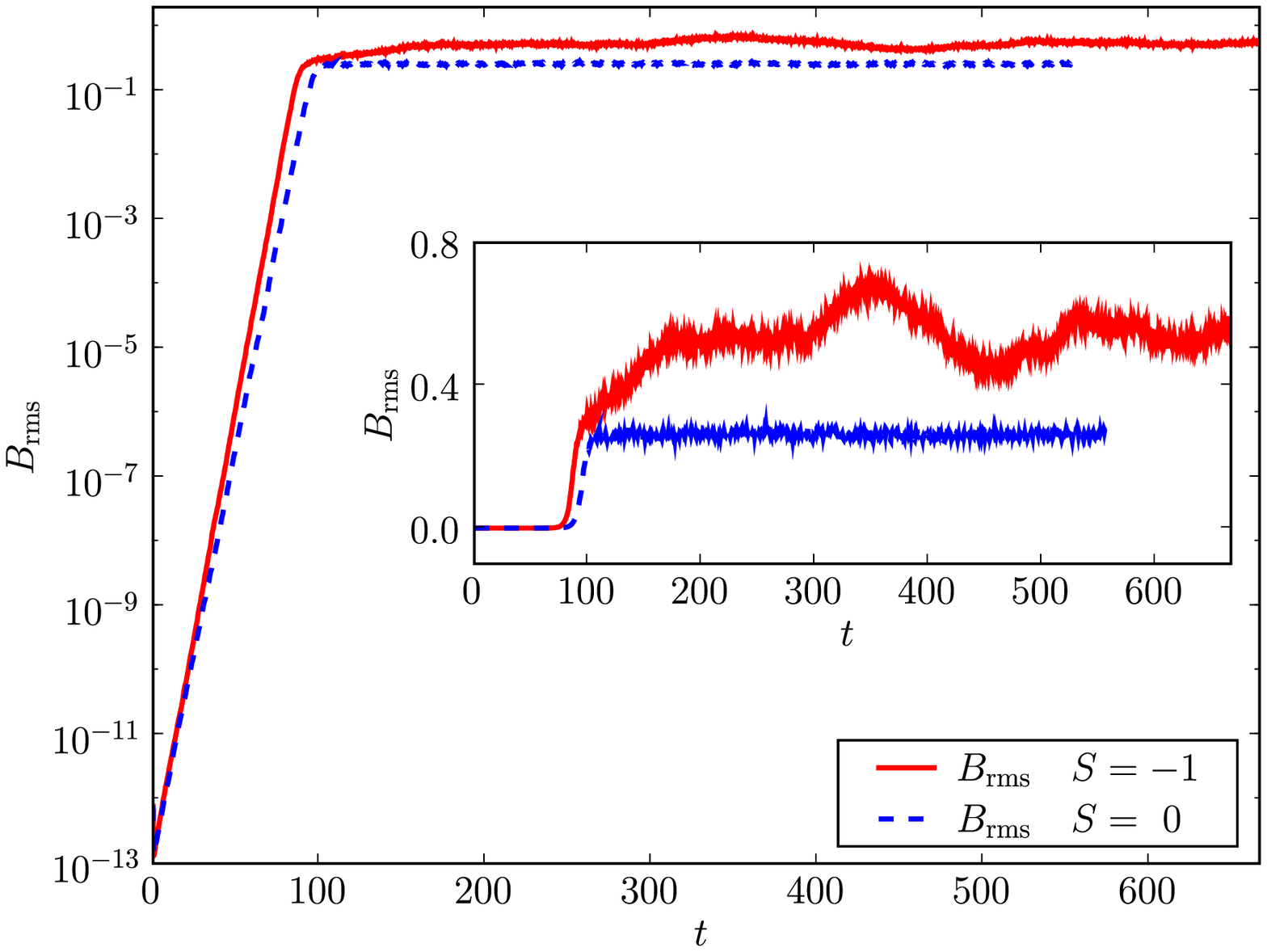}\quad
\includegraphics[width=.49\textwidth]{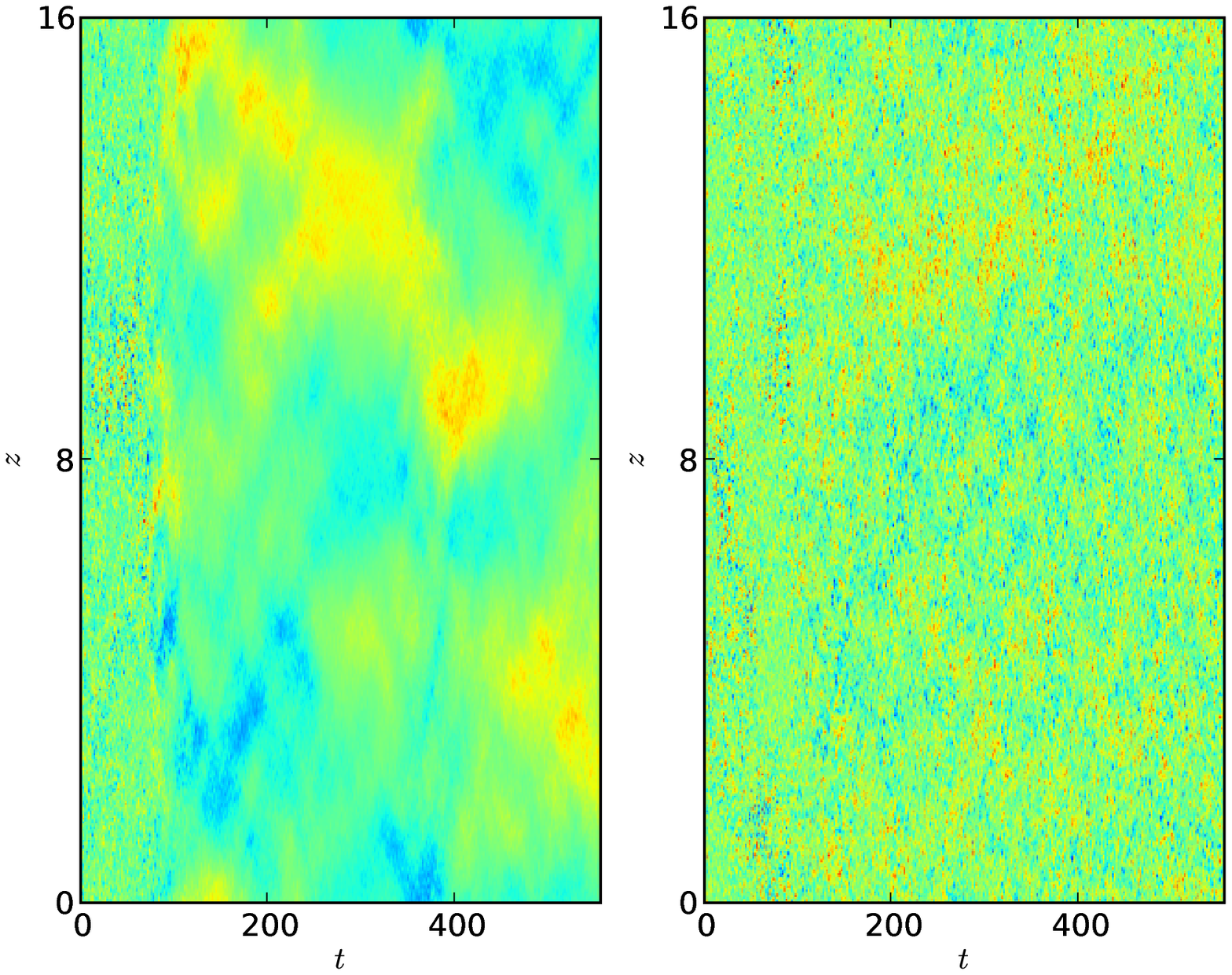} 
\caption{(online colour at: www.an-journal.org) \emph{Left panel}: time evolution of $\Brms$ for runs FD1  ($S=-1$ and
$\Omega=2/3$) and FD2 ($S=\Omega=0$). 
The inset shows the same evolution on a linear scale.
\emph{Right panel}: The $y$ component of the magnetic field averaged over $x$ and $y$,
$\xyavg{B_y}(z,t)$, for runs FD1 (left) and FD2 (right). 
The run FD1 develops large-scale structure with long correlation time, while
the run FD2 only develops small short-lived fluctuations.}
\label{fig:fd}
\end{figure*}

\begin{figure}
\includegraphics[width=.49\textwidth]{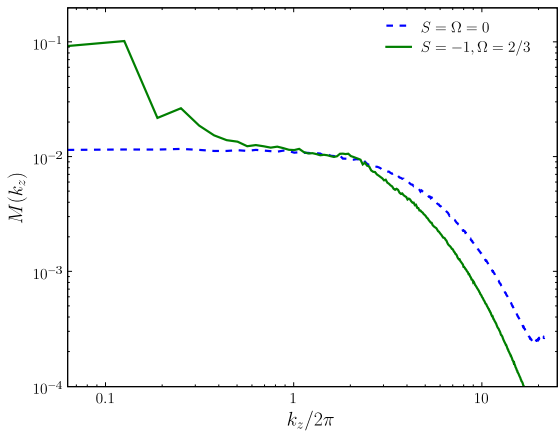}
\caption{(online colour at: www.an-journal.org) Time averaged normalised  one-dimensional magnetic energy spectra
(Eq. 3) for FD1 ($S=-1$, $\Omega=2/3$) and FD2 ($S=\Omega=0$). 
The time average is taken over $t=200$ to $500$.  
Unlike for FD2, the spectrum for FD1  peaks at low wave numbers (corresponding
to large scales). 
This indicates the existence of large-scale magnetic structure.}   
\label{fig:fdspec}
\end{figure}

So far, we have only discussed simulations with $\Rm$ below the critical value for
the onset of the fluctuation dynamo $\Rmc$.
This choice was made to ensure that the growth in magnetic energy is 
solely due to the shear dynamo, allowing us to confirm the existence of this
effect, its independence of the fluctuation dynamo and to study it in a
``pure'' isolated form. 
However, astrophysical objects virtually always have large $\Rm$, 
which are expected to be far above the fluctuation dynamo threshold. 
Thus, any mean-field dynamo scheme aspiring to explain large-scale 
astrophysical fields has to work simultaneously with the 
fluctuation dynamo. Furthermore, the fluctuation 
dynamo is likely to be much faster than any mean-field dynamo 
because its growth rate is determined by the turbulent 
rate of strain associated with scales much smaller than 
the outer scale, so $\gamma\gg\tau^{-1}$, where 
$\tau$ is the outer-scale turbulent turnover time 
\citep[see discussion in][and references therein]{sch07}. 
In contrast, mean-field dynamos are typically slower 
than the outer-scale rate of strain, so for them, 
$\gamma\ll\tau^{-1}$. Thus, a working mechanism for mean-field 
generation must operate against the background not of pure 
hydrodynamic turbulence but rather of magnetohydrodynamic 
turbulence that emerges as the saturated state 
of the fluctuation dynamo \citep[the approach advocated, e.g., 
by][]{raedler03,RK03,RK04}. 
It cannot be taken for granted that such a magnetohydrodynamic 
turbulence in conjunction with shear will still be germane 
to the generation of large-scale magnetic fields. 
It is, therefore, imperative to confirm the existence 
of the shear dynamo effect for $\Rm>\Rmc$ in order to be able to
assert its potential relevance for astrophysical situations. 
Here we show (as far as we know, 
for the first time) that a combination of linear shear 
and a nonhelically forced homogeneous {\em magnetohydrodynamic} 
turbulence (saturated state of fluctuation dynamo) can support 
dynamically significant large-scale magnetic fields. 

For this preliminary study, we performed two runs, FD1 and FD2. 
For FD1, $S=-1$, $\Omega=2/3$; for FD2, $S=\Omega=0$. 
Both runs have $\nu=\eta=1\times10^{-3}$, $\urms\sim1$ and thus $\Re=\Rm\approx50$. 
This value is above the threshold for the fluctuation dynamo. 
In the run FD2, which is a pure homogeneous turbulence run, 
that is the only possible field-amplification mechanism. 
In contrast, run FD1 has the prerequisites for both the shear and fluctuation
dynamos. 

The time evolution of the magnetic field for these two runs
is shown in \figref{fig:fd}. Both runs develop in a
similar manner initially: the magnetic field  is amplified by the
fluctuation dynamo and $\Brms$ increases by 12 orders of magnitude to 
a dynamically strong saturated level after a time $t\sim 100 \sim 100\tau$. 
This growth is much faster than the growth rates associated with the 
shear dynamo in analogous runs without the fluctuation dynamo.
The growth rate is $\gamma\simeq 0.3$, which 
is consistent with the expectation that the growth rate 
of the fluctuation dynamo should be comparable to the turbulent 
rate of strain, or stretching rate---for $\Rm\ge\Re$, 
this is usually the rate of strain associated with the smallest 
turbulence scales \citep[see, e.g.,][and references therein]{sch04,sch07,brand05review}, 
but in our case the scale separation between the 
outer and the viscous scale is not large, so the fluctuation-dynamo 
growth rate should be roughly comparable to $\tau^{-1}$. 

As evident in \figref{fig:fd}, the two runs differ significantly after the
saturation of the fluctuation dynamo. 
The magnetic field for the zero-shear run FD2 becomes statistically stationary
and stays at small scales,  while in the sheared and rotating run FD1 it
develops large-scale structure. 
The structural difference between the runs with and without shear is displayed
in the right panel of \figref{fig:fd}: whereas the zero-shear run only has
small scale, short-lived turbulent fluctuations, the run with an imposed shear
develops large-scale, long-lived structures, which are in many ways similar to
the shear-dynamo-generated mean field in the runs without the fluctuation
dynamo.
The existence of large-scale structures in the case with shear can also be seen
from (time-averaged normalised one-dimensional)  magnetic energy spectra shown
in \figref{fig:fdspec}.
The spectrum for FD1 ($S=-1$,$\Omega=2/3$), unlike that for FD2 ($S=\Omega=0$),
has a prominent peak at a low wave numbers corresponding to a large-scale mean
field.
Note that $\Brms$ in FD1 appears to develop long-lived, large-amplitude, 
large-scale fluctuations that are reminiscent of those observed
in $\urms$ for the vorticity dynamo reported in \secref{sec:nlinst}
(see the inset in the left panel of \figref{fig:fd}). 

Thus, we have found that the fluctuation dynamo and the shear dynamo 
can coexist, giving rise both to small- and large-scale magnetic fields. 
It is clear that a systematic parameter study at high $\Rm$ and $\Re$ is called for. 
It requires much larger computational 
resources than the relatively cheap simulations reported in this paper. 
Since the seed field for generating large-scale magnetic fields 
is now provided by the already saturated small-scale fields 
resulting from the fluctuation dynamo, it is never really well 
posed under these circumstances to consider a kinematic regime of 
the shear dynamo \citep[cf.][]{cattaneo08}. 
Thus, we have to confront the question of how the large-scale 
fields saturate and what their structure is in a fully nonlinear dynamical 
regime. We have left these questions outside the scope of this paper. 
Further investigations along these lines will be reported elsewhere. 

\section{Conclusion}
\label{sec:conc}

Can nonhelical turbulence in combination with large-scale velocity shear act
as a mean-field dynamo and generate magnetic fields with length scales much
larger than the outer scale of the turbulence? 
This has been the subject of considerable recent 
debate \citep{vishniac97,urpin99a,urpin99b,silantev00,urpin02,fedotov03,RK03,RK04,brand05,fedotov06,raedler06,ruediger06,raedler07,proctor07,brand07,kleeorin08}. 
To our knowledge, the paper by \citet{yousef08} and this
paper present the first set of dedicated numerical experiments that
demonstrates that such a generation mechanism is feasible. 
However, in retrospect, one might conjecture that the shear-dynamo  
might have already been seen in several earlier numerical
studies that combined large-scale flows, and consequently large-scale shear,
with nonhelical forcing at smaller (or, in some cases, similar) scales 
and reported generation of magnetic fields at scales larger 
than the forcing scale \citep{brand05,ponty05,mininni05,shapovalov06}. 

We have carried out a suite of numerical experiments on the shear 
dynamo effect in vertically elongated 
shearing boxes and for magnetic Reynolds numbers
subcritical with respect to the fluctuation dynamo. 
For the values of the imposed shear $S$ between $1/8$ and $8$ 
(corresponding to $S\tau\sim 0.04 \dots 3$, where $\tau$ is the 
turnover rate of the randomly forced velocity fluctuations), 
we have found that the dynamo growth rate is $\gamma\propto S$ 
and the characteristic length scale of the generated mean magnetic field 
is $\lb\propto 1/\sqrt{S}$. 

The first key result of this paper, compared 
with the earlier study by \citet{yousef08}, is that the shear dynamo 
works both in the nonrotating case and for the case of Keplerian 
rotation ($\Omega=-2S/3$). There does not appear to be much difference, qualitative 
or quantitative, between the rotating and nonrotating cases, although 
perhaps it would be interesting to look at non-Keplerian cases 
and try to identify the role of rotation via a parameter scan in 
$\Omega$ independent of the one in $S$.  

The second key result, claimed on the basis of only a preliminary study, 
is that the shear dynamo works both for situations that are sub- and supercritical
with respect to the fluctuation (small-scale) dynamo. In the latter case, 
the overall magnetic energy grows very quickly due to the fluctuation 
dynamo effect independent of the presence of the shear. Imposing the shear 
on the magnetohydrodynamic turbulence resulting from the saturation 
of the fluctuation dynamo, leads to the emergence of magnetic fields 
that have spatial scales 
larger than the outer scale of the turbulence and that fluctuate 
on very long time scales compared to the turbulent turnover time.  

Thus, the shear dynamo effect appears to be quite general and robust. 
As the combination of a shear flow and turbulence is a very common
feature in astrophysical systems, the shear dynamo potentially
represents a generic mechanism for making large-scale fields. While
much needs to be understood about its properties before its relevance 
to real astrophysical systems can be more than an appealing speculation,
the simplicity of the idea of the shear dynamo certainly makes it a
worthwhile object of study. It is also important to determine how generic 
the shear dynamo is and how it combines with other large-scale 
features present in real astrophysical systems: various differential 
rotation laws, temperature and density gradients, linear instabilities, etc.. 

Studies in this vein are already being undertaken. 
For example, recent numerical experiments by \citet{kapyla08} have shown that 
large-scale dynamo action is also possible in local simulations of
magnetoconvection with imposed horizontal shear. 
They also report a growth rate $\gamma \sim S$ and find large-scale 
dynamo action for magnetic Reynolds numbers above the critical threshold
for the fluctuation dynamo. Another topical recent study is by \citet{gressel08}, 
who simulated the supernova-driven galactic turbulence and found that 
they needed to impose a linear velocity shear to obtain the amplification 
of a large-scale field. These studies clearly demonstrate the key role of 
shear in producing a mean-field dynamo. However, in comparing their 
results to ours, one has to keep in mind 
that their simulations had rotation and vertical stratification, 
so the turbulence in these simulations is likely to be helical and may also host 
an $\alpha$ effect. 

In motivating our choice of Keplerian rotation law, we mentioned the 
possible relevance to accretion-disk turbulence, which is believed to be 
driven by the magnetorotational instability (MRI) \citep{balbus03}. 
Given a (weak) large-scale field, the MRI will generate velocity 
and magnetic-field fluctuations at small scales. 
These fluctuations, in conjunction with Keplerian rotation and shear, 
must then amplify the large-scale field to close the loop. 
The mechanisms for such an ``MRI dynamo'' have been discussed 
and simulated for some time
\citep[e.g.,][]{brand95,hawley96,fromang07,rincon07b,rincon08,lesur08}. 
 It is tempting to observe in the context of the results reported 
above that a combination of small-scale turbulence (magnetohydrodynamic 
turbulence in the case of the MRI) and large-scale shear does indeed 
appear to work as a dynamo giving rise to a large-scale azimuthal 
magnetic field ($B_y$ in the shearing sheet model). We note, however, 
that \citet{lesur08}, who have analysed this process in detail, 
find some important differences between what happens in MRI-driven shearing sheet 
simulations and the forced case studied by us. 

To conclude, we believe that the discovery of shear dynamo has opened 
a number of new and exciting avenues of research and produced some 
promising leads towards unravelling the ways in which cosmic magnetic 
fields emerge. Further investigations will help assess the range of applicability
and relevance of the shear dynamo effect and the physical mechanisms
that are responsible for it.

\acknowledgements
We are grateful to E.~Blackman, A.~Brandenburg, L.~Kitchatinov, G.~Lesur, G.~Ogilvie,
J.~Papaloizou, S.~Fromang, D.~Shapovalov and D. Sokoloff for important discussions.
TAY, NK and IR thank Nordita for its hospitality during the \emph{Turbulence and Dynamos} 
program and the participants of the program for discussions of the shear dynamo effect.
The numerical simulations were performed on HPCF (Cambridge), DCSC (Copenhagen),
NCSA (Illinois) and UKAFF.
This work was supported by the UK STFC (TAY, TH and AAS) and 
Isaac Newton Trust (TAY and TH). 
We also acknowledge travel support from the Royal Society (IR), 
US DOE Centre for Multiscale Plasma Dynamics (TAY, TH, FR and AAS)
and the Leverhulme Trust Network for Magnetized Plasma Turbulence (TH and FR).

\end{document}